# Droplet Formation via Solvent Shifting in a Microfluidic Device

Ramin Hajian[1,2,a], Steffen Hardt[1,b]

[1] Center of Smart Interfaces, Technical University of Darmstadt, Alarich-Weiss-Straße 10, 64287 Darmstadt, Germany.

[2] Department of Mechanical Engineering, Iranian University of Science & Technology, Narmak, 1684613114, Tehran, Iran.

**Abstract**:

"Solvent shifting" is a process in which a non-solvent is added to a solvent/solute mixture and extracts the solvent. The solvent and the non-solvent are miscible. Because of solution supersaturation a portion of the solute transforms to droplets. In this paper, based on this process, we present an investigation on droplet formation and their radial motion in a microfluidic device in which a jet is injected in a co-flowing liquid stream. Thanks to the laminar flow, the microfluidic setup enables studying diffusion mass transfer in radial direction and obtaining well-defined concentration distributions. Such profiles together with Ternary Phase Diagram (TPD) give detailed information about the conditions for droplet formation condition as well as their radial migration in the channel. The ternary system is composed of ethanol (solvent), de-ionized water (non-solvent) and divinyle benzene (solute). We employ analytical/numerical solutions of the diffusion equation to obtain concentration profiles of the components. We show that in the system under study droplets are formed in a region of the phase diagram between the binodal and the spinodal, i.e. via a thermally activated process. The droplets are driven to the channel centerline by the solutal Marangoni effect but are not able to significantly penetrate into the single-phase region, where they get rapidly dissolved. Therefore, the radial motion of the binodal surface carries the droplets to the centerline where they get collected.

## I.  Introduction

Generating microdroplets in microfluidic devices has been in the focus of intense research efforts over the last decade[1,2]. Such an interest is related to the manifold applications of droplet microflows. For example, they may be used for the production of chemical compounds or nanoparticles[3-6], for the encapsulation of biological cells[7,8] and other biological studies[9], or for generating tailor-made emulsions with monodisperse droplets or for materials with monodisperse particles.[10-13]

Conventional methods of droplet production in microfluidic devices are mainly based on T-junctions, flow focusing[1,2] or 3D axisymmetric co-flow droplet generators.[14-16] In these approaches the droplet formation is either governed by hydrodynamic drag forces or by surface-tension driven hydrodynamic instabilities, most notably the Rayleigh-Plateau instability. By varying channel dimensions, flow rates and even temperature one can produce droplets of different sizes. However, it is obvious that for a given microfluidic design with a given dimension of the nozzle or other geometric features, the droplet size will always be closely linked to these scales. While it is possible to produce smaller droplets by decreasing the channels size, in practice it is often not convenient to prepare and work with microfluidic chips in which the geometric dimensions are only a few micrometers. These facts boost motivations to think about methods for droplet generation in which the droplets are much smaller than the geometric dimensions.

---

[a] Email: hajian@csi.tu-darmstadt.de , raminhajian@yahoo.com
[b] Email: hardt@csi.tu-darmstadt.de (The corresponding author)

Mary et al.[17] studied in detail the extraction process between droplets and the continuous phase in microchannels. Yu et al.[18] also investigated such extraction processes in a microfluidic chip. They visualized the extraction by fluorescent imaging and presented a relationship between the extraction efficiency and the droplet size. Sang et al.[19] investigated the droplet formation process by extraction of a gas into liquid. They injected microbubbles composed of perfluorohexane and carbon dioxide into water. As $CO_2$ is soluble in water but $C_6F_{14}$ is not, only $CO_2$ was extracted from the bubbles and the bubbles shrank. Reduction of the bubble size led to a pressure increase, and finally a transition into the liquid state, i.e. formation of $C_6F_{14}$ micro-/nanodroplets.

An alternative way of producing small droplets is via a solvent shifting process. A hydrophobic substance such as anise oil can be dissolved in a solvent such as ethanol, giving a transparent solution. By adding a third component like water, which is miscible with the solvent but not with the oil, the solvent (i.e. ethanol) mixes with the third component (i.e. water), leaving behind oil dissolved in a liquid with an increasing concentration of the third component. From a certain point on the solution becomes supersaturated, and eventually small oil droplets nucleate. Because these droplets scatter light, the appearance of the liquid becomes opaque. This phenomenon has been known for a long time but, to the best of our knowledge, has only been investigated in detail since the 1970's. Ruschak and Miller[20] investigated this phenomenon in a water-ethanol-toluene system, theoretically and experimentally. Vitale and Katz[21] used divinyl benzene (DVB) as oil, ethanol as solvent and de-ionized (DI) water as the third component. They provided a ternary phase diagram (TPD) for this system which in addition to single-phase and two-phase regions also shows the region in which small stable droplets form by homogeneous nucleation. Vitale and Katz termed this phenomenon "ouzo effect" because people observe it by adding water to ouzo, a drink which contains ethanol and anise oil. Grillo[22] studied this classical system using small-angle neutron scattering and measured sizes of oil droplets. Sitnikova et al.[23] studied the trans-anethol-water-ethanol system, specifically the anethol droplet growth, using the dynamic light scattering. They found that the evolution of the droplet size distribution can be explained by Ostwald ripening. They also affirmed that there is no coalescence among droplets. Such a spontaneous emulsification process is based on "solvent shifting"[24]; we prefer to use this term here because it does not refer to any specific ternary system and seems more general. Some researchers took advantage of the solvent shifting effect (SSE) to produce very small particles.[24-26] McCracken and Datyner[26] studied the water-methanol-styrene system and proposed the SSE as a method for the production of submicron-sized particles. Hung et al.[27] followed the solvent shifting approach to synthesize micro-/nanoparticles in microchannels. They first generated bigger microdroplets and then by extracting the solvent from droplets to the surrounding phase, supersaturation occurs inside the droplets, and micro-/nanoparticles form. Karnik et al.[28] generated nanoparticles via the SSE and nanoprecipitation processes in a flow focusing microfluidic device.

In none of the previous efforts on the SSE the evolution of the concentration profiles, resulting in supersaturation and droplet formation, has been studied. Usually, the mixing process going along with the SSE is not well controlled, so information on how droplet formation is related to the local concentrations of the three components is not available. The whole process – from nucleation to Ostwald ripening – could be influenced by imposing well-defined concentration profiles, an option that does not seem to have been explored so far. Nevertheless, microfluidic techniques were identified as powerful candidates to investigate the mechanisms acting in the early stages of the SSE.[29] In the present article, based on the laminar flow in the microfluidic device, we study the SSE using well-defined concentration profiles. These profiles guide us to determine the local composition of the system at the point where droplets are forming. Moreover, the corresponding microfluidic device also allows us to study the dynamics following droplet nucleation, which would be impossible based on the common process using rather large vessels. An important point in that context is that however tiny droplets are generated by the SSE that are potentially stable in a homogeneous mixture,[21] the existence of a concentration gradient in the continuous phase induces a motion of these droplets driven by Marangoni stresses. That way droplets can accumulate in certain regions and can possibly coalesce to form bigger droplets. Understanding the mechanism of droplet formation and their subsequent dynamics in the microfluidic device could be an initial step to control the droplet size distribution.

In this article, we present an experimental/theoretical investigation on mass transfer and droplet formation in co-flowing streams in which droplets are formed via the SSE. The fluid of the inner stream is a mixture of ethanol and DVB, and that of the outer stream flow is de-ionized water (DI water). Quite close to the nozzle DVB droplets form and migrate to the channel centerline via a complex process which we identify as the interplay of Marangoni convection and the inward motion of the binodal surface (*BS*). The result is an accumulation of the produced droplets at the center of the channel.

## II. Experimental

### A. Setup

Figure 1 shows a schematic of the experimental setup. The microfluidic device is composed of a tapered round glass capillary (nozzle) inside a square glass capillary (denoted as channel in the following). This geometry allows a fluid flow which is approximately axisymmetric in the vicinity of the centerline of the channel. The inner and outer diameters of the nozzle are 30 μm and 50 μm, respectively. The width of the channel is about 1 mm. Because of the considerable difference in fluid densities, shown in Table I, the inner fluid stream which mainly contains ethanol will deviate to the top wall of the channel if the microfluidic device is arranged horizontally. To prevent such an undesired disturbance of the flow, the channel is oriented vertically. The fluids are injected into the microfluidic device using syringe pumps (KD Scientific 210). A long distance microscope (INFINITY K2/SC, objective CF-4), and a high-speed camera (MotionPro Y4) are used for imaging.

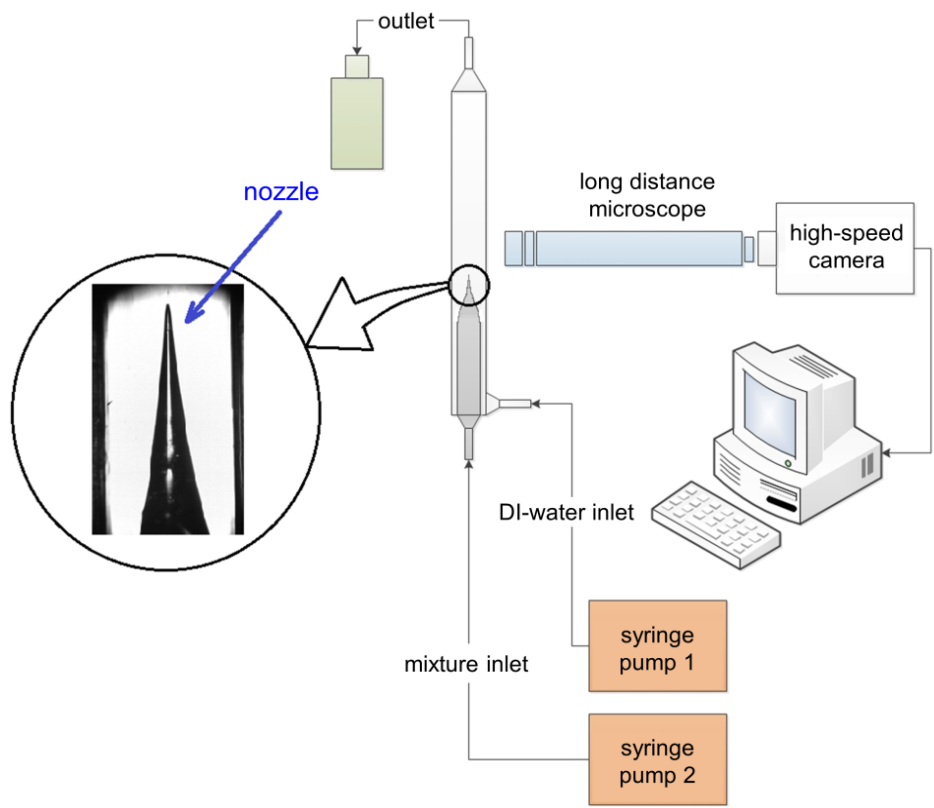

FIG. 1. Schematic of the experimental setup

| Component | Density (kg/m$^3$) | Viscosity (mPa·s) |
|---|---|---|
| DVB | 914$^{(1)}$ | 1$^{(2)}$ |
| ethanol | 785$^{(3)}$ | 1.1$^{(3)}$ |
| DI water | 1000$^{(3)}$ | 0.9$^{(3)}$ |

TABLE I. Material properties (at 25 °C): (1) Online data for Sigma-Aldrich product: DVB technical grade 80%; (2) DVB Product Stewardship Manual, Dow Chemical Company, USA (2000); (3) D.R. Lide, *CRC Handbook of Chemistry and Physics*, 89$^{th}$ ed. 2008-2009.

**B. Materials**

The mixture in the inner stream (jet) is composed of 99.5 weight percent (wt%) ethanol and 0.5wt% divinyl benzene (DVB). Because of the low concentration of DVB, the purity of ethanol must be as high as possible otherwise the concentration of DVB would be close to that of the impurities, with unknown consequences for the phase diagram of the mixture of liquids which could no longer be considered ternary. Therefore, we used LiChrosolv gradient grade ethanol with a purity of more than 99.9%, provided by Merck Millipore. DVB of 80% technical grade was used, provided by Sigma-Aldrich. It contains some polymerization inhibitor and other impurities. Nonetheless, we used it as received from the manufacturer since after removing the polymerization inhibitor, DVB will not last as a monomer for a long time and must be used up soon because its properties change with time, which would have been impractical for the series of experiments conducted. However, to assure that purified and unpurified DVB result in the same TPD, we reproduced some points of the published phase diagram[21] and observed that the results are very similar. In addition, we used an inhibitor remover column, which was provided by Sigma-Aldrich, to remove the polymerization inhibitor of DVB, i.e. we purified DVB. Then we repeated some of the experiments with the microfluidic device and observed that the results are the same as those obtained with unpurified DVB. Hence, usage of unpurified DVB for the experiments and analysis according to the TPD presented by Vitale and Katz[21] is acceptable without considerable errors.

### C. Experimental conditions

We performed all experiments at lab temperature (22-23 ºC). The flow rates for the inner (mixture) and outer (water) stream are 3 and 50 microliters per minute (μl/min), respectively (abbreviated as 50/3). Figure 2 shows the jet that is formed under these conditions. Actually the nozzle direction is vertical (as seen in Figure 1), therefore the jet also flows in vertical upward direction, however in Figure 2 it is rotated clockwise by 90°. The *jet region* is indicated by *d*, being the jet diameter. In this paper, we analyze droplets formation and migration at the specified flow rates. However, for reasons we discuss in subsection IV.C we also did experiments with other flow rates - i.e. ranging from 50/2 to 50/10 - to vary the jet diameter.

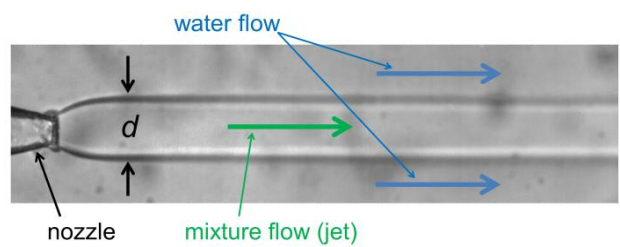

FIG. 2. Jet of the liquid mixture flowing parallel to an outer stream of water (jet flow rate = 3 μl/min, water flow rate = 50 μl/min).

### III.  Mass transfer analysis

Since ethanol and water are miscible, there will be a diffusive mass flux of ethanol into the outer stream, accompanied by a diffusive mass flux of water into the jet. According to Table I the viscosity inside the channel is almost constant (and approximately equal to $10^{-3}$ Pa·s). On the other hand, the jet development length, i.e. the distance from the nozzle at which the jet diameter reaches its final value, is small, e.g. less than 80 μm for the 50/3 flow rate. Thus, the fluid velocity profile over the whole channel cross section is close to that of fully developed laminar flow of a fluid with constant viscosity in a square channel. Typical Reynolds numbers are of the order of one, showing that it is justified to assume laminar flow. In order to analyze mass transfer in the radial direction, for simplicity we assume a uniform flow velocity in the *diffusion region*, i.e. a cylindrical region within the central part of the channel in which diffusion mass transfer takes place.

We define the *diffusion time* as the time it takes until the whole jet region, which is initially in the single-phase state, transforms into a two-phase fluid. As we explain in section IV, the relevant *diffusion time* in the case of 50/3 is less than 2 s. As mentioned above, the lateral diffusion of jet fluid (mainly ethanol) into the surrounding fluid (DI water) occurs simultaneously with the diffusion of DI water into the jet region. Therefore, during the *diffusion time* the *diffusion region* also broadens owing to the smearing of the concentration profiles. For a time scale of 2 s the corresponding length scale is about 40 μm. By adding that to the jet radius (i.e. 50 μm for the case of 50/3), the whole *diffusion region* has a radius of 90 μm. According to this approximation the *diffusion region* extends over about 20% of the channel width. The velocity gradients are minimal in the central part of the channel. Therefore, assuming a uniform velocity in this region is not expected to cause significant errors in the mass transfer analysis. That way the advection-diffusion problem is reduced to a pure diffusion problem. Furthermore, in the central part of the (square) channel the deviations from an axisymmetric situation will still be small. We can hypothetically study the concentration profiles in the whole channel cross section as the diffusion zone and apply boundary conditions at the channel wall. However, velocity gradients existing in the outer regions of the channel cross section will not invalidate the simplified picture employed here if we only focus at the *diffusion region* over a time span of the order of the *diffusion time*. To solve the diffusion equation we have to apply boundary conditions at the channel wall.

Therefore, we can solve the diffusion equation in a cross-sectional plane co-moving with the flow velocity at the center of the channel to determine the concentrations of the three components. Such a model does not account for diffusion in axial direction. For the flow rates considered the axial concentrations gradients are very small compared to the gradients in radial direction. Therefore, only accounting for radial diffusion is a valid approximation. We solve the diffusion equation for two components (i.e. ethanol and DVB) to find their concentrations, i.e. their weight ratios, and then determine the concentration of the third one (i.e. water) by subtracting the contributions of those two components from unity. The transient diffusion equation for the axisymmetric case is given by

$$\frac{\partial C}{\partial t} = \frac{1}{r}\frac{\partial}{\partial r}(rD\frac{\partial C}{\partial r}), \qquad (1)$$

where $D$ and $C$ are the diffusion coefficient and concentration of the relevant component, respectively, and $r$ is the radial distance from the channel centerline. Since - as mentioned before - the viscosity is approximately constant across the channel, the Stokes-Einstein relation relating viscous dissipation and diffusion suggests that $D$ can be regarded as a constant. The solution of Eq. (1) for an initial concentration distribution of $f(r)$ and the outer surface (i.e. the channel wall) being impermeable is[30]

$$C(r,t) = \frac{2}{a^2}\left\{\int_0^a xf(x)dx + \sum_{n=1}^{\infty}\exp(-D\alpha_n^2 t)\frac{J_0(r\alpha_n)}{J_0^2(a\alpha_n)}\int_0^a xf(x)J_0(\alpha_n x)dx\right\}, \qquad (2)$$

where $J_i$ are Bessel functions of the first kind, $\alpha_n$ are roots of $J_1(a\alpha_n)$ and $a$ is half of the channel width, i.e. 0.5 mm. The initial conditions are

$$f_{eth}(r) = \begin{cases} 0.995 & r \leq \frac{d}{2} \\ 0 & r > \frac{d}{2} \end{cases} \qquad f_{DVB}(r) = \begin{cases} 0.005 & r \leq \frac{d}{2} \\ 0 & r > \frac{d}{2} \end{cases},$$

where $d$ is the diameter of the jet. The channel shape does not play a role in Eq. (2) because the *diffusion region* is far away from the walls.

Eq. (2) will be valid as long as the system is in the single-phase region, or as long as the disturbance of the system caused by the formation of DVB droplets is negligible. When diffusive mass transfer proceeds, phase change occurs and droplets of DVB form. The number density of these droplets is low and also the size of droplets is smaller than 1 μm (see section IV). So, due to the sparseness of droplets we assume that DVB droplet formation does not interfere considerably with ethanol diffusion, and Eq. (2) should be valid to describe the latter. $D_{eth}$ is approximately equal to $8.4\times10^{-10}$ m$^2$/s, which is the diffusion coefficient of ethanol in water.[32] The presence of DVB changes this coefficient very little, since its viscosity is close to that of the other components. Concentration profiles for ethanol are presented in the next section. According to the Stokes-Einstein equation, at constant viscosity the diffusion coefficient is inversely proportional to the size of molecules. Considering this and knowing $D_{eth}$ we estimate $D_{DVB}$ to be $6.2\times10^{-10}$ m$^2$/s.

The TPD presented for the ethanol-DVB-water system by Vitale and Katz[21] was used to produce the diagram shown in Figure 3. In that work, the grey region was denoted *stable ouzo region*, which shows the compositions at which DVB droplets form. Also they termed *unstable ouzo region* the region where phase decomposition occurs. In the *stable ouzo region* the concentration of DVB is very low; when the supersaturation in the system is sufficiently high, DVB nuclei form - through a thermally activated process - with high nucleation rates. Each nucleus depletes its surrounding regions of DVB molecules while it is growing and the droplet is forming; this is a fast process occurring on a scale of milliseconds or faster.[21] The droplets are so distant from each other that they hardly coalesce and stay stable for a long time without using surfactants. This is not the case in the *unstable ouzo region*; in this region the DVB concentration is high enough or that of ethanol is sufficiently small that the usual spinodal decomposition characterized by a vanishing or very small free-energy barrier for nucleation occurs.

If we use Eq. (2) also to determine the concentration of DVB, a typical concentration curve at a given time - in the following denoted as *diffusion curve* - will be of the form of the dashed curve in Figure 3. A *diffusion curve* is a trajectory in concentration space, at a certain time, with increasing values of the radial position, starting at the channel centerline and ending at the wall. It hits the binodal curve (see Figure 3) and enters the *stable ouzo region* which leads to formation of very small droplets of DVB. At $t = 0$ the radius of the jet marks the radial position of the *stable ouzo region,* i.e. in that case it is only a point in the space of radial coordinates. At later stages, owing to diffusional broadening, this region extends to a finite interval in *r*. This region extends continuously so that finally the whole jet region is covered.

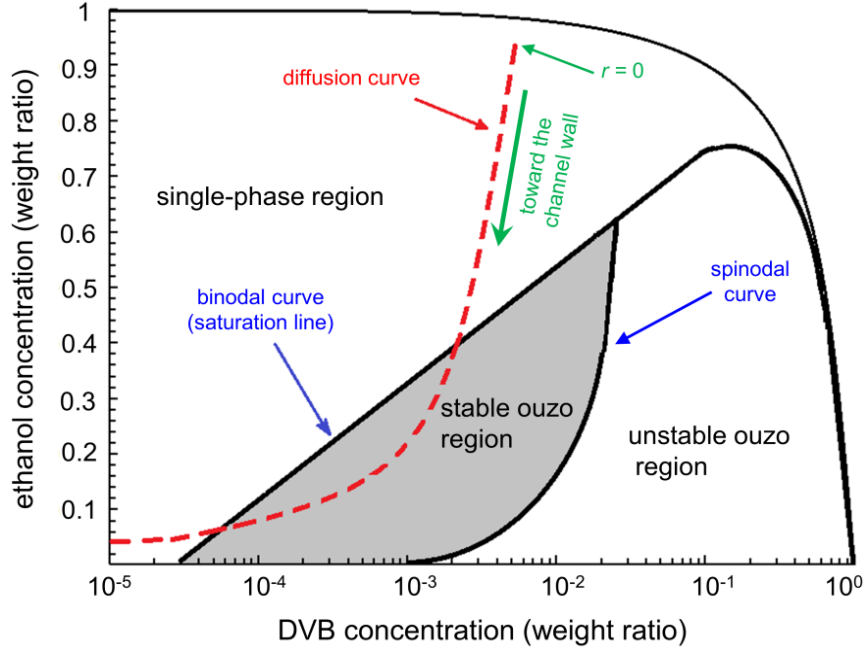

FIG. 3. Ethanol-DVB-water ternary phase diagram (TPD) with a typical *diffusion curve*.

So far, however, we have neglected one effect in the diffusion equation for DVB that will change this picture to some extent. Formation of DVB droplets means that we need a sink term in the corresponding diffusion equation, representing the transition of dissolved molecules into droplets. The sink term represents an average over a region containing many droplets, in the spirit of a homogenization scheme. Owing to the sparseness of the DVB droplets, they do not significantly reduce the spatial domain in which diffusion can occur. Thus, the modified diffusion equation for DVB is

$$\frac{\partial C_{DVB}}{\partial t} = \frac{1}{r}\frac{\partial}{\partial r}(rD_{DVB}\frac{\partial C_{DVB}}{\partial r}) - ST, \quad (3)$$

where *ST* is the sink term. We can solve this equation numerically if we know *ST* as a function of $C_{DVB}$, $t$ and $r$. The sink term is the result of DVB droplet formation, which occurs via nucleation if the solution is sufficiently supersaturated.[21] The amount of DVB per unit time and unit volume converted into droplets depends on the size distribution and number density of droplets present at a given time, which in turn depends on the history of nucleation events. A corresponding model was presented by Hasan et al.[33] in the form of

$$\Gamma_G(t) = \int_0^t \left\{ 4\pi r'^2 \frac{\partial r'(t,t')}{\partial t} \rho_{DVB} \right\} J(t')dt', \quad (4)$$

where $\Gamma_G(t)$ is the time dependent rate of DVB mass per unit volume transferred from the dissolved phase to droplets. $r'$ is the radius of a set of droplets at time $t$ with their nuclei formed simultaneously at a former time $t'$. The sizes of the nuclei of each set are the same, and they grow with an identical rate. $\rho_{DVB}$ is the mass density of DVB. $J(t')$ is the nucleation rate (i.e. the number of nuclei formed per unit time and unit volume) of critical clusters at time $t'$ and is expressed as

$$J = A\exp(-\frac{\Delta G^*}{k_B T}), \qquad (5)$$

where $\Delta G^*$ is the free energy needed to form a critical cluster. $k_B$ and $T$ are the Boltzmann constant and the absolute temperature, respectively. The prefactor $A$, $\Delta G^*$, and $r'$ depend on both $C_{DVB}$ and the concentration of DVB at saturation ($C_{DVB,sat}$). $C_{DVB}$ and $C_{DVB,sat}$ are both time dependent.

Using Eq. (4) as a sink term in the diffusion equation for DVB would transform this equation into a nonlinear integro-differential equation, i.e. a highly complex mathematical framework. Furthermore, for quantitative modeling of the sink term one would have to determine the parameters entering Eq. (5), which is far from trivial. At the current stage we are aiming at studying the principal effects of droplet formation on the DVB concentration field rather than at a quantitative description. Therefore, rather than relying on eqs.4 and 5 we presently use a simpler linear model for the sink term. Subsequently, in subsection IV.C we discuss how much the existence of a sink term affects the results. In fact, there we compare the solutions of the diffusion equation for DVB with and without sink term and show that the existence of a sink term does affect the main conclusions of our analysis. The linear alternative for the sink term is determined as follows. We define the supersaturation as

$$SS = \frac{C_{DVB} - C_{DVB,sat}}{C_{DVB,sat}}, \qquad (6)$$

We assume that the nucleation is triggered at a certain supersaturation – let us call it the maximum supersaturation ($SS_{max}$). The respective concentration of DVB is $C_{DVB,max} = (1 + SS_{max}) \times C_{DVB,sat}$. We also assume that as soon as $C_{DVB}$ exceeds $C_{DVB,max}$, the excess concentration which is equal to ($C_{DVB}$ - $C_{DVB,max}$), transforms into droplets very fast. The supersaturation is not necessarily constant for various compositions; however, for the moment we assume a hypothetical value of $SS_{max}$ equal to 10%. At the end of section IV we show that even by large changes of $SS_{max}$ the main results are not much affected.

Based on these arguments we rewrite Eq. (3) as

$$\frac{\partial C_{DVB}}{\partial t} = \frac{1}{r}\frac{\partial}{\partial r}(rD_{DVB}\frac{\partial C_{DVB}}{\partial r}) - k(C_{DVB} - C_{DVB,\max}), \qquad (7)$$

where $k$ is a kinetic constant representing the speed of droplet growth. Results obtained for the same ternary system indicate that it takes a very short time from the onset of nucleation until a DVB droplet reaches its final size.[21,31] Therefore, here we explore the limit that droplet formation as an infinitely fast process. Consequently, $k$ should be so large that further increasing it does not change the DVB concentration profiles considerably. We tried $k = 1$ s$^{-1}$, 10 s$^{-1}$, 100 s$^{-1}$, etc., and found this condition is satisfied if $k = 10000$ s$^{-1}$, i.e. for $k > 10000$ s$^{-1}$ the results do not change significantly. If $C_{DVB} < C_{DVB,max}$, the supersaturation is not high enough to trigger nucleation and form droplets so the sink term is omitted (i.e. $ST = 0$). As mentioned above, this kind of modeling of the sink term is not necessarily accurate, but helps to highlight the principal effects of DVB droplet formation.

Eq. (7) can be numerically solved if we know $C_{DVB,sat}$ as a function of $t$ and $r$. As seen in Figure 3, the section of the binodal curve at the border of the *stable ouzo region* is approximately linear, termed *saturation line*; the relevant line equation is easily specified. Any arbitrary point on this line represents a saturation condition. As a result, the corresponding concentrations of DVB and ethanol, on the abscissa and ordinate of the diagram, are $C_{DVB,sat}$ and $C_{eth,sat}$, respectively - $C_{eth,sat}$ is the ethanol concentration at saturation. At certain $t$ and $r$, if we know the respective point on the *saturation line*, we can find the related $C_{DVB,sat}$. On the other hand, using Eq. (2), the ethanol concentration is known at a given $t$ and $r$. $C_{eth}$ is obtained independently of the solution state (i.e. subsaturated, saturated or supersaturated). Therefore, for given $t$ and $r$, we can use the computed value of $C_{eth}$ at this space-time point to identify the point at the saturation line that gives the same value of $C_{eth}$. The DVB concentration obtained from that point is identical to $C_{DVB,sat}$.

## IV.   Results and discussions

### A. Concentration distributions

Figures 4(a)-(c) show concentration distributions of the various components in terms of the weight ratio as a function of radial position and time. The origin of the $x$ axis is located at the center of the channel. The radial domain is limited to a maximum of $r = 200$ μm because after that point concentrations are approximately constant. Figure 4(a) displays the ethanol concentration resulting from the exact solution, i.e.

Eq. (2). Figure 4(b) shows the DVB concentration that is obtained from the numerical solution of Eq. (7). The concentration of DI water, presented in Figure 4(c), was obtained by subtracting that of ethanol and DVB from unity at each $r$ and $t$. Initially, the concentrations of DVB and ethanol are maximal in the *jet region* (extending from the origin to $r = 50$ μm) and equal to zero in the rest of the domain. As diffusive mass transfer takes place, the two-phase region, initially located at $r = 50$ μm, expands both toward smaller and larger values of $r$. As shown in Figure 4(d) the boundaries of the expanding domain can be computed based on the analytically and numerically obtained concentration fields; the ethanol concentration fields are displayed as contours. The upper white and the lower black curves represent the outer and the inner phase separation boundary, respectively. The region between these two curves is in the two-phase state and the rest, i.e. below the black curve and above the white one, is in the single-phase state. Figure 4(d) also shows that at 1.5 s the center of the channel transforms from a single-phase into a two-phase region.

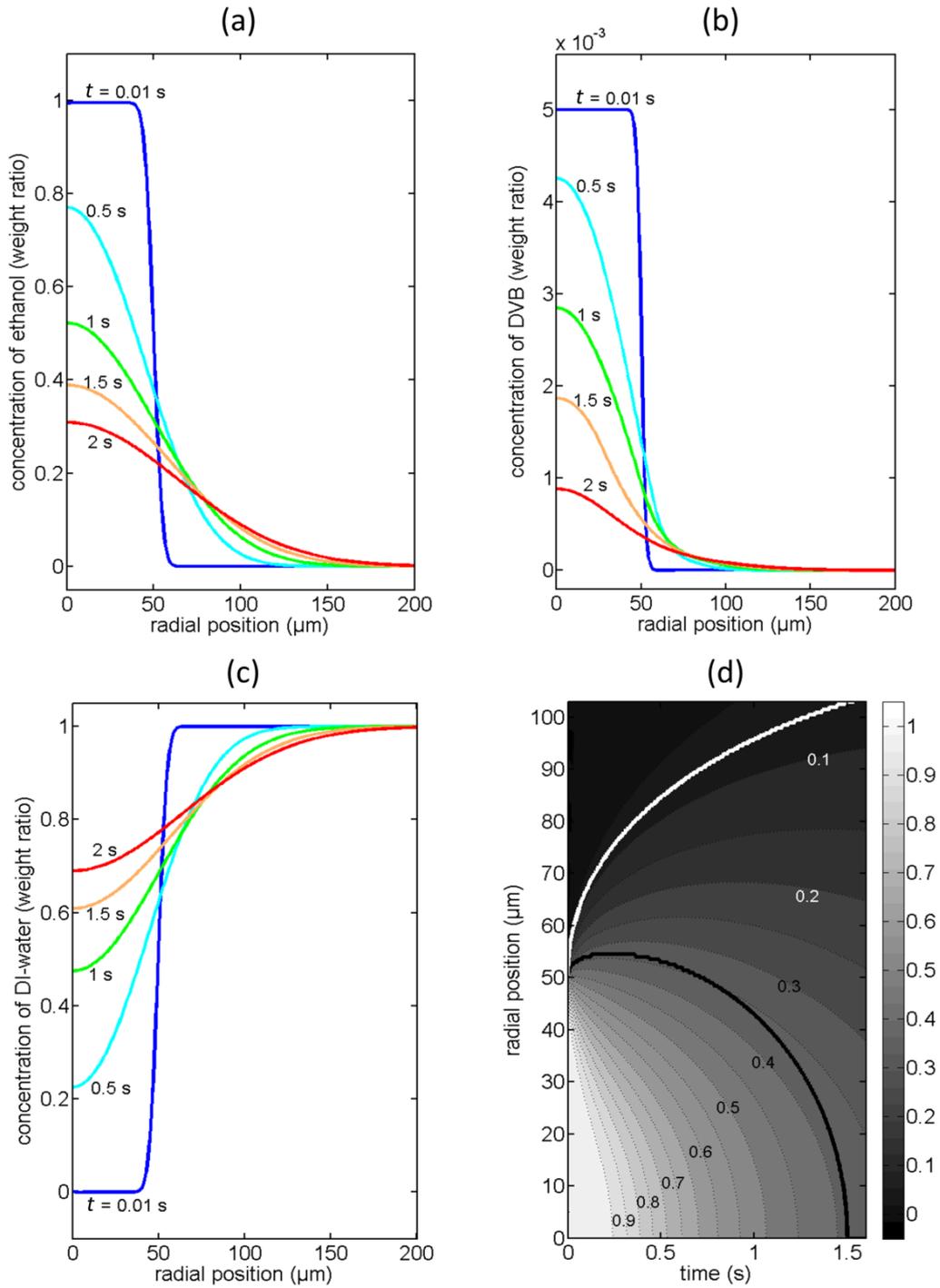

FIG. 4. Concentration profiles of (a) ethanol, (b) DVB (with sink term) and (c) DI water as a function of time for the experiment with the 50/3 flow rate ratio. (d) Contours of ethanol concentration and boundaries between single-phase and two-phase regions. The white and the black curves show the outer and inner boundary, respectively.

## B. Droplet formation

Plotting *diffusion curves* in the TPD (Figure 5) we find the points at which they cut the binodal curve. These points determine the radial positions $r$ where phase separation occurs. By comparing Figure 3 and Figure 5 we discover that the *diffusion curves* hit the binodal curve and enter the *stable ouzo region*, i.e. the region where DVB droplets form. The schematic *diffusion curve* which is shown in Figure 3 does not account for the sink term. Similarly, the top of Figure 5 shows numerically computed diffusion curves without sink term. When applying the sink term in the DVB diffusion equation, i.e. Eq. (7), those parts of the *diffusion curves* lying within the two-phase region become shifted toward the *saturation line* (Figure 5 bottom), because a portion of dissolved DVB is extracted into droplets, resulting in a decrease of the DVB concentration. All of the diffusion curves start at $r = 0$. The fact that the starting points of these curves move closer and closer to the binodal as time progresses indicates that larger and larger portions of the radial domain transform into a two-phase region.

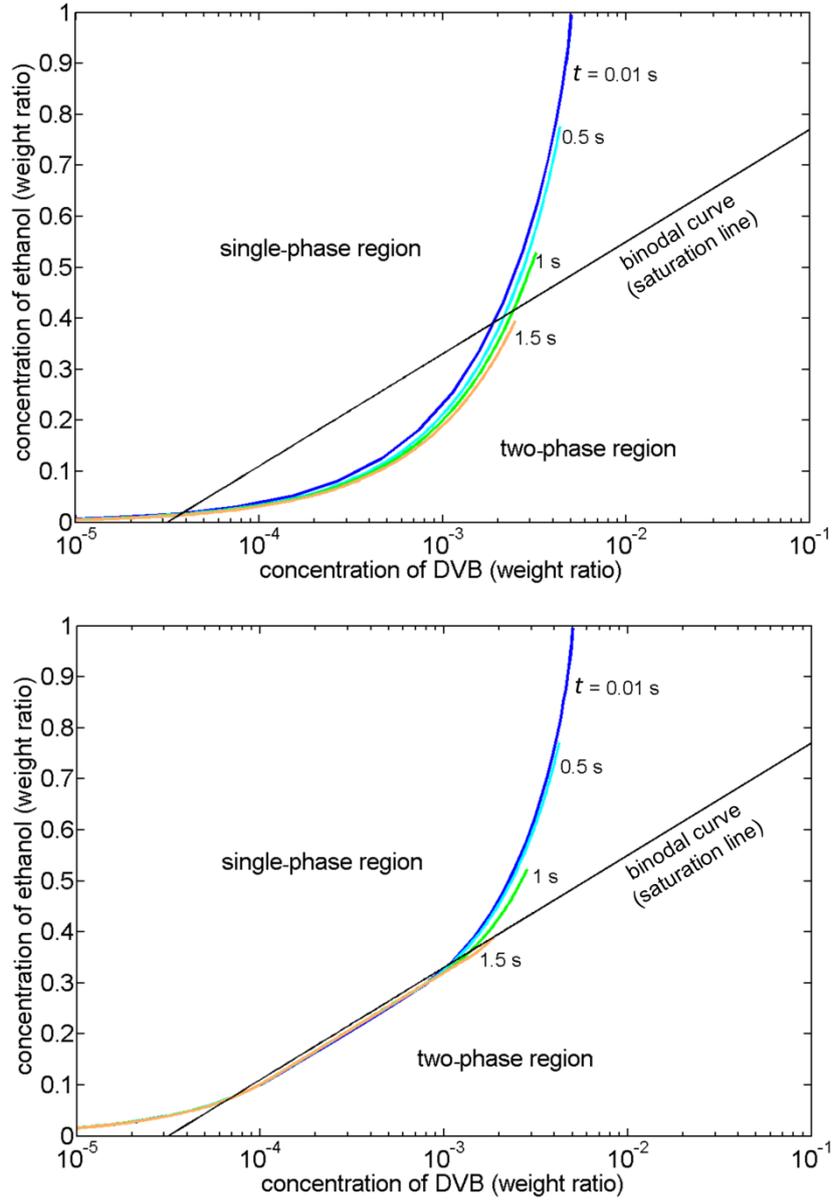

FIG. 5. *Diffusion curves* at three different points in time intersecting with the binodal curve of the TPD without (top) and with (bottom) sink term in the diffusion equation for DVB.

According to the literature,[21,31] the size of the droplets in the *stable ouzo region* varies between 0.8 and 3 μm. This result was obtained by experiments in which all of the components were filled in a vessel with certain mass ratios. In such experiments both convection and diffusion contribute to the mass transfer. This is not the case for the experiments of the present study in which mass transfer in radial direction occurs merely via diffusion. Since the mixing protocol may affect the droplet size distribution it can *a priori* not be expected that the experiments with the microfluidic device yield similar droplet sizes as reported in the above mentioned references.

The experiments with the microfluidic device were performed using a long-distance microscope (INFINITY K2/SC, with CF-4 objective). However, we temporarily used another objective (Nikon 20X objective + Navitar 2X adapter) to maximize the optical resolution. We also took videos with a very high frame rates (up to 10000 frames per second). Even with such an optical/imaging enhancement single droplets could not be identified. By contrast, with the same setup test particles of 2 μm diameter in a suspension were clearly visible. Based on that, we estimate that the DVB droplets created in the microfluidic device are at least one order of magnitude smaller (i.e. of the order of 100 nm). Because of difficulties in adjusting the focal plane such an enhanced optical setup could not be used permanently, so the main imaging was performed with the long-distance microscope and the CF-4 objective. With this combination the droplets in the microchannel appear in the form of black streaks when their number density is high enough (see the next section). As we discuss in detail in subsection IV.C, the droplets get collected at a certain radial position as time proceeds. This radial distance decreases with time, such that the droplets finally accumulate at the centerline of the channel.

**C. Radial droplet migration**

As soon as DVB droplets are formed, they move radially toward the center of the channel. Figure 6 shows an example of black streaks composed of small DVB droplets. The figure shows edited images taken at different axial positions along the channel. In fact, in the raw images the streaks are often not clear enough. So the editing, performed on images corresponding to $z = 1$ mm to $z = 4$ mm, was done in such a way that lines were drawn to better highlight the position and orientation of the streaks. The frame at the upper right is a raw image which shows streaks of droplets almost at the center of the channel. In the experiments with the 50/3 flow rate ratio, they arrive at the channel centerline somewhere between $z = 4$ mm and $z = 5$ mm. The result is a stream of small droplets arranged along the centerline. The figure also shows the boundary of the jet ($z = 0$) which is clearly visible close to the nozzle and becomes increasingly fuzzy away from the nozzle, owing to water and ethanol diffusing into each other.

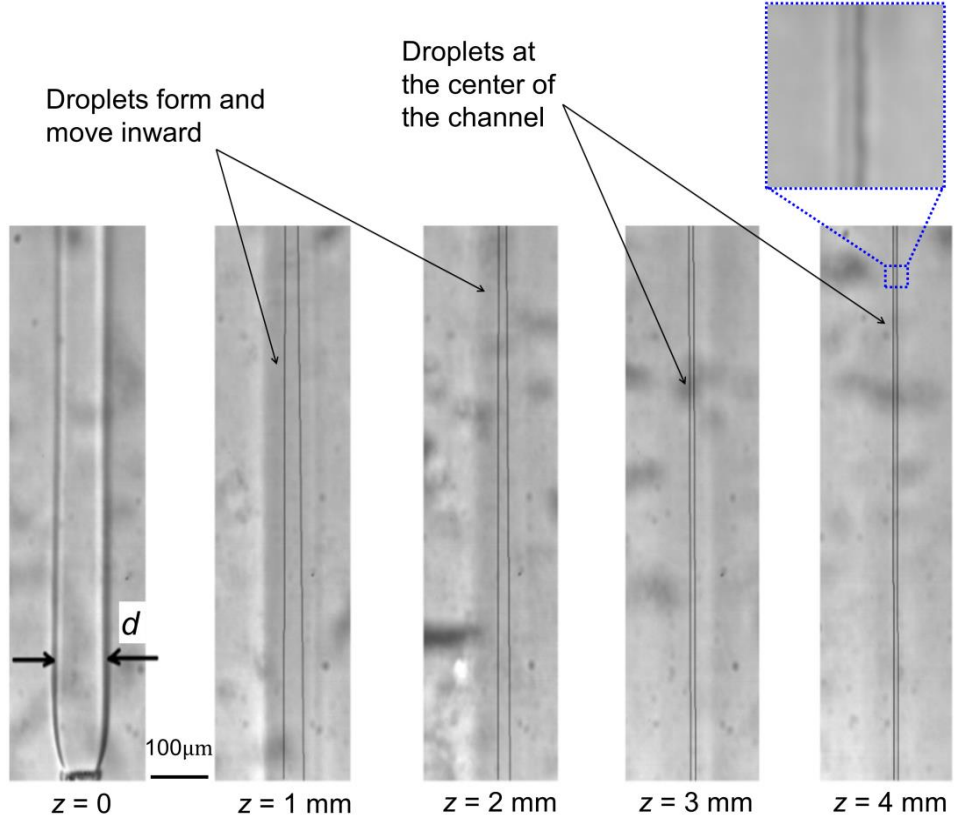

FIG. 6. Formation of DVB droplets and their radial migration in the experiment with the 50/3 flow rate ratio. From $z =$ 1 mm to $z = 4$ mm streaks are highlighted to become better visible. The frame corresponding to $z = 0$ and the frame at the upper right show raw images.

To understand the reason for such radial migration of droplets, we briefly review the phenomenon of droplet migration driven by Marangoni stresses. If a droplet is immersed in another liquid with more than one component, the concentration gradient in the surrounding liquid usually causes a gradient in the interfacial tension at the droplet surface. Such a gradient in the interfacial tension leads to the solutal Marangoni effect. This means that a flow along the interface is generated, pointing from the region of low to the region of high interfacial tension. As a result, the fluid around the droplet flows in the direction of the interfacial tension gradient, while the droplet moves in the opposite direction. This phenomenon is similar to thermocapillary motion, which has been studied mainly since the 1950s for both bubbles[34,35] and droplets.[35-39] There are also some published results related to the solutal Marangoni effect for components with no surface activity (i.e. no surfactants). [40,41] One can estimate the order of magnitude of the droplet velocity from the tangential stress balance at the interface,[37,38] giving

$$u \sim \frac{R_0 G_C \sigma_C}{\mu}, \qquad (8)$$

where $u$ is the droplet velocity, $R_0$ is the droplet radius, $\mu$ is the dynamic viscosity of the continuous phase. Also, $\sigma_C = \partial\sigma/\partial C$ represents the variation of the interfacial tension ($\sigma$) with concentration, and $G_C = \partial C/\partial r$ is the concentration gradient at the position of the droplet. The concentration of DVB is very small compared to that of ethanol and DI water. Therefore, also the DVB concentration gradients are expected to be small compared to those of the other components. For this reason either the water or the ethanol concentration field can be used to compute the Marangoni stresses, while the DVB concentration gradients will be neglected. Here the ethanol concentration is used for this purpose.

As soon as DVB droplets form, they are exposed to concentration gradients of the components. The concentration of ethanol at the side of the droplet closer to the channel centerline is more than that on the other side; for water the opposite is true. DVB is soluble in ethanol but not in water, so it is expected that the interfacial tension between the DVB droplets and surrounding liquid decrease with increasing concentration of ethanol. Therefore, the interfacial tension should decrease toward the center of the channel, and due to the solutal Marangoni effect droplets should move into the same direction (see Figure 7(a)).

To confirm the hypothesis about the direction of the interfacial tension gradient, we measured the interfacial tension for different weight ratios of ethanol and water. To obtain the value of $\sigma_C$ we used a Profile Analysis Tensiometer (model PAT1, provided by SINTERFACE Technologies) to measure the interfacial tension between a DVB droplet and mixtures of ethanol and DI water. The measurement method is based on the shape analysis of a buoyant droplet. The results are presented in Figure 7(b). As seen in this figure, the interfacial tension between DVB and the mixtures decrease with increasing concentration of ethanol, as expected. The data also show that the variation of interfacial tension with concentration of ethanol is approximately linear, so $\sigma_C$ is constant and takes a value of approximately 87 (with the unit of mN/m per ethanol weight ratio). The data in Figure 7(b) were obtained by averaging five readings for each experiment, and the error bars represent the standard deviations for each set of readings.

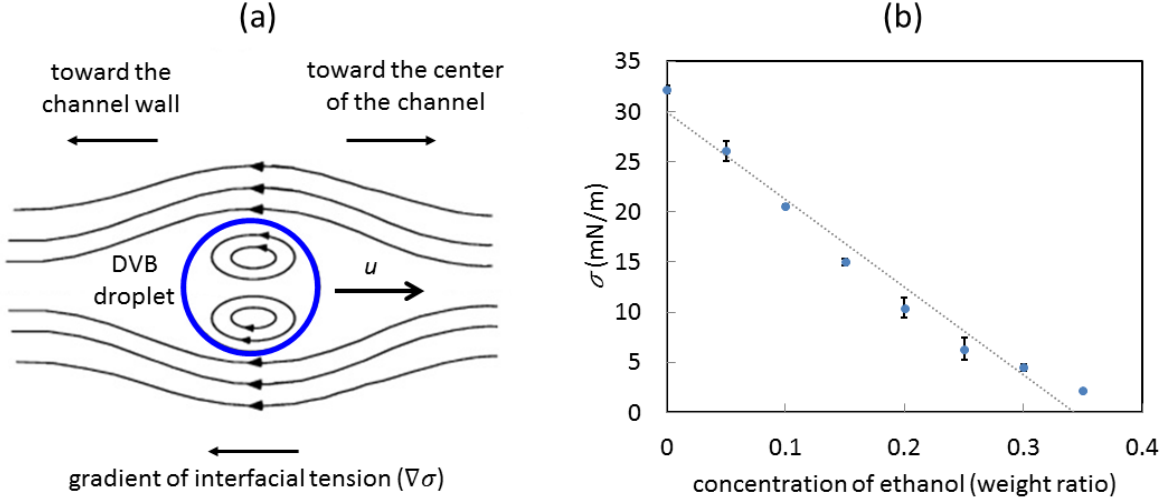

FIG. 7. (a) Schematic of the radial migration of a droplet driven by the solutal Marangoni effect. (b) Variation of the interfacial tension between DVB and ethanol/DI water mixtures, obtained by buoyant droplet interfacial tension measurements.

Based on the above discussion and using Eq. (8) we are now in a position to estimate the radial droplet velocity due to Marangoni convection and compare it with experimental results. Even if Eq. (8) does not give an exact value for the droplet velocity, it is useful for finding its order of magnitude. For estimating that velocity we assume a droplet diameter of 0.1 μm and a viscosity of 1 mPa·s. While the value of the droplet diameter is based on the reasoning presented in subsection IV.B, the viscosity value is very close to what is known for the system under study. From the solution of the respective diffusion equation the concentrations of water and ethanol are known at each radial position and point in time, which provides $G_C = G_C(r,t)$ The DVB droplets are so small that inertia effects can be neglected. The equation for the radial motion of a droplet is then given as

$$\frac{dr}{dt} = u(r,t), \qquad (9)$$

where $u(r,t)$ is determined from Eq. (8). Eq. (9) can be integrated to determine the droplet travel time from $r = d/2$ to $r = 0$:

$$\Delta t = \int dt = \int_{d/2}^{0} \frac{dr}{u(r,t)}. \qquad (10)$$

The average radial velocity of the droplet can be calculated through $\bar{u} = 2\Delta t / d$. In the case of the 50/3 flow rate ratio, $\bar{u}$ and $\Delta t$ for a droplet diameter of 0.1 μm are 329 μm/s and 152 ms, respectively. However, the experimentally measured travel time of the droplets (i.e. $t_{drop}$, explained in the following) is about 2.1 s, i.e. more than one order of magnitude bigger than what is predicted by the Marangoni effect. In fact it turns out that solely assuming radial droplet transport by the Marangoni effect is an oversimplified picture that needs to be complemented, as will be discussed below.

From the instant at which the first droplets form it takes some time until the whole *jet region* transforms from the single-phase to the two-phase state (see subsection IV.A, especially Figure 4(d)). Due to the solutal Marangoni effect the droplets move toward the center of the channel. As long as a droplet is still located at the two-phase region it does not lose mass because its surrounding is at a saturated (or even a non-fully relaxed supersaturated) condition. As soon as it enters the single-phase region the droplet starts to dissolve. The DVB droplet loses mass continuously while moving toward the channel centerline. The time it takes the droplet to fully dissolve gives an idea about whether or not it is able to reach a specific radial position (e.g. the center of the channel).

For that purpose, we assume a simple scenario in which the droplet is stationary at a point with a certain concentration in its surroundings. Under these conditions the rate of mass transfer can be estimated via[42,43]

$$-\frac{dm}{dt} = 4\pi R D_{DVB} (\rho'_{sat} - \rho'_{sur}) = 4\pi R D_{DVB} \Delta \rho' , \qquad (11)$$

where $m$ is the mass of the droplet, $R$ is the radius of the droplet, and $\rho'_{sat}$ and $\rho'_{sur}$ are the partial densities of DVB at saturation condition and in the surrounding fluid, respectively. $\rho'_{sat}$ depends on the droplet radius through the Kelvin equation:[44]

$$\frac{\bar{R} T}{M_{DVB}} \ln \frac{S}{S_\infty} = \frac{2\sigma}{\rho_{DVB} R} , \qquad (12)$$

where $\bar{R}$ is the *universal* gas constant, $T$ is the absolute temperature, and $M_{DVB}$ is the molar mass of DVB. $S$ is the saturation concentration of DVB in the saturation layer around the droplet. $S_\infty$ is the saturation concentration of DVB on a plain surface which is the normal solubility (i.e. equal to $C_{DVB,sat}$). $\rho_{DVB}$ is the density of DVB. The interfacial tension ($\sigma$) between the DVB droplet and the surrounding liquid is of the order of 1 mN/m. That is so small that for the estimation of the droplet lifetime we can safely assume $S = S_\infty$, i.e. the saturation concentration is independent of the droplet size. Therefore the only variable on the right hand side of Eq. (11) is $R$. Now the differential equation can be solved by expressing the droplet mass through its radius:

$$-\frac{dm}{dt} = -\frac{d(\rho_{DVB}V_{drop})}{dt} = -4\pi R^2 \rho_{DVB}\frac{dR}{dt},$$

where $V_{drop}$ is the *volume* of the droplet. From Eq. (11) we get

$$-4\pi R^2 \rho_{DVB}\frac{dR}{dt} = 4\pi R D_{DVB}\Delta\rho',$$

*resulting* in

$$\Delta t_{dis} = \frac{\rho_{DVB}R_0^2}{2D_{DVB}\Delta\rho'}, \qquad (13)$$

where $\Delta t_{dis}$ is the time it takes until the droplet is fully dissolved and $R_0$ is the initial droplet radius. Assuming that the *surrounding* liquid is composed of ~ 40 wt% ethanol, ~ 60 wt% DI water and ~ 0.15 wt% DVB and $R_0 \approx 50$ nm, $\Delta t_{dis}$ is approximately equal to 2 ms. This is a very small time scale compared to the droplet travel time (i.e. 152 ms), estimated based on the Marangoni effect. The supposed chemical composition corresponds to the single-phase region close to the binodal line (see Figure 5 bottom). Therefore, as soon as a droplet crosses the *BS* and enters the single-phase region, it gets dissolved very fast. The *BS* is the cylindrical surface which is the boundary between single-phase and two-phase regions; its position at a given time is determined via the black curve in Figure 4(d). It is moving toward the center of the channel, so the already dissolved DVB transforms to droplet(s) again when it encounters the *BS*. In other words, for the assumed droplet size there is a periodic process of droplet formation, transport, dissolution and repeated formation. Droplets that are formed well within the two-phase region are carried to the *BS* by Marangoni convection, but effectively are unable to cross it. The periodic process of droplet formation and fast dissolution results in an accumulation of droplets at the *BS*. This explains the appearance of inward-moving black streaks, as visible in Figure 6.

Owing to the resolution limits of the imaging system, the size of the order of 100 nm is an upper bound rather than a reliable size estimate. For the case that the DVB droplets are even smaller than that, $\Delta t_{dis}$ decreases very fast (~ $R_0^2$), while also the droplet speed due to the Marangoni effect decreases (~ $R_0$). In any case, one is left with the conclusion that while the droplets being formed are transported toward the channel centerline, a scale analysis reveals that as soon as they cross the *BS* they are dissolved so fast that they are not able to penetrate into the single-phase region to any significant degree. Therefore, we suggest that the inward motion of the *BS* carries the droplets to the channel centerline.

To evaluate this hypothesis we study the correlation between the travel time of the *BS* ($t_{bs}$) to the center, which is found from the solution of the diffusion equation, and that of droplets ($t_{drop}$), determined experimentally by measuring the time the black streaks take to reach the centerline. Theoretically they should be identical, however, because of some errors discussed at the end of this section they are not expected to coincide exactly

As seen in Figure 6, for the experiment with the 50/3 flow rate ratio, droplets, represented by black streaks, reach the center of the channel somewhere between $z = 3.5$ mm and $z = 4.5$ mm. The error in determining the axial position where the droplets reach the centerline is estimated to be about 0.5 mm. The average velocity ($V_{ave}$) of the fluid in the channel is known via the total flow rate. The processes of droplet formation and radial migration happen near the channel centerline, so the velocity in the *diffusion region* is close to the maximum velocity. Utilizing the velocity profile in a square channel,[45] the maximum velocity ($V_{max}$) is calculated as 2.1 times the average velocity ($V_{ave}$). Knowing $z$ together with the flow velocity, $t_{drop}$ is obtained. As shown in Figure 4(d), for the 50/3 case it takes 1.5 s until the *BS* reaches the center of the channel, i.e. $t_{bs} = 1.5$ s.

By changing the flow rate of the jet and consequently the jet diameter, the lateral travel distance of droplets (which is equal to $d/2$) changes and we obtain different values of $t_{bs}$ and $t_{drop}$. Experiments were carried out for a constant water flow rate of 50 μl/min and various jet flow rates of 2 to 10 μl/min; $t_{drop}$ was measured for each flow rate in the way described above. Also, the diffusion equation was solved for all of these flow rates and corresponding $t_{bs}$ values were obtained. The solutions were performed for the diffusion equation with and without the sink term; the results presented in Figure 8 include both cases. In this figure, each point represents $t_{bs}$ and $t_{drop}$ for a certain jet flow rate. By increasing the flow rate, both $t_{bs}$ and $t_{drop}$ increase. For instance, in Figure 8 the first triangle/square on the left correspond to the 50/2 flow rate ratio, while the last symbols correspond to 50/10. This figure also demonstrates that to a good approximation, $t_{bs}$ and $t_{drop}$ are linearly correlated.

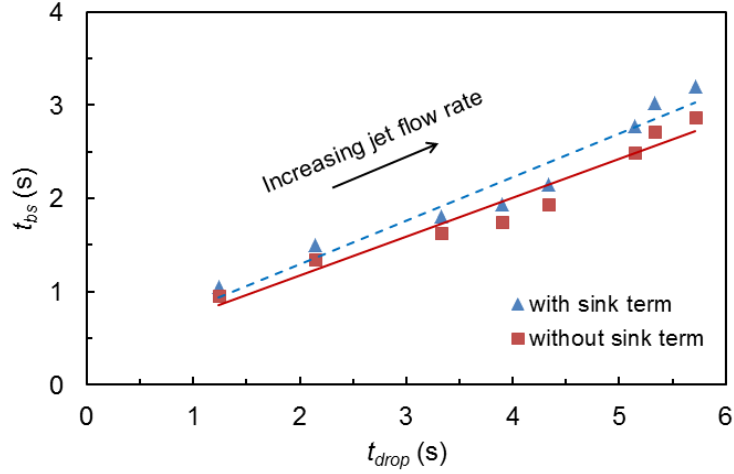

FIG. 8. Comparison between the measured travel time of droplets ($t_{drop}$) and the calculated travel time of the *BS* ($t_{bs}$) for both cases: with and without sink term in diffusion equation.

As displayed in Figure 8, $t_{bs}$ and $t_{drop}$ are close to each other and follow the same trend. This supports the hypothesis formulated above, i.e. that the droplets are carried to the channel centerline with the *BS*. In Figure 8 results for two different cases are shown, corresponding to the diffusion equation for DVB with and without sink term. The correlations between $t_{bs}$ and $t_{drop}$ for these two cases are quite similar. Bearing in mind that a simplified description for the sink term was chosen, this indicates that the details of DVB mass transfer to droplets are irrelevant in the present context. Actually, the existence of a sink term does alter the basic conclusion about the correlation between $t_{bs}$ and $t_{drop}$.

The results presented correspond to $SS_{max} = 10\%$. For lower and higher supersaturations the same conclusions can be drawn about the correlation between $t_{bs}$ and $t_{drop}$. We solved Eq. (7) not only for $SS_{max} = 10\%$ but also for other values of $SS_{max}$ and observed no significant changes in the results for $t_{bs}$. For instance, for the 50/3 flow rate ratio with $SS_{max} = 10\%$, $t_{bs}$ is equal to 1.5 s. The value shifts to 1.54 s and 1.47 s if $SS_{max}$ is changed to 0.1% and 20%, respectively. This means that using a rather arbitrary value for $SS_{max}$ (such as 10%) is an acceptable approach.

If our hypothesis was correct, $t_{bs}$ and $t_{drop}$ would have to be identical for a specific jet flow rate. The difference between $t_{bs}$ and $t_{drop}$, observable in Figure 8, may come from diverse sources. The diffusion coefficient of DVB in a water/ethanol mixture or in ethanol has not been found in the literature, so we needed to estimate it based on the Stokes-Einstein equation, which may have introduced errors in determining $t_{bs}$. Also, diffusion gets inhibited in regions with a high density of DVB droplets, i.e. close to the *BS*. An important aspect not to be overlooked is that the determination of the ternary phase diagram shown in Figure 3 is challenging in itself. Especially identifying the binodal and spinodal with adequate accuracy is a difficult task. Last but not least, measurement errors in the determination of $t_{drop}$ can also play a role in that regard.

## V.   Conclusion and future work

A combined experimental/analytical study was accomplished on the solvent shifting process in a microfluidic device with 3D axisymmetric co-flow configuration. Solvent shifting is a process in which a non-solvent such as water extracts a solvent such as ethanol from a solvent/solute mixture. As a result, a portion of the solute appears as droplets; that way the system relaxes from supersaturation. The laminar character of the flow inside the channel made it possible to study the mass transfer in the radial direction in detail. Small submicron DVB droplets form in the two-phase region of the flow domain in a thermally activated process. These are visible as dark streaks that move radially inward and finally form a streak of DVB droplets at the channel centerline. The experimental results were compared with analytical and numerical solutions of the radial diffusion equation, allowing to predict the evolution of the binodal surface (*BS*). It was found that the formed DVB droplets move radially inward driven by the solutal Marangoni effect. However, the droplets are virtually unable to cross the *BS* and penetrate significantly into the single-phase region where they rapidly dissolve. Therefore, it can be concluded that the motion of the *BS* toward the channel centerline is responsible for the accumulation of DVB droplets at the center. This was confirmed by comparing the experimentally recorded time scale for droplet transport with the theoretical prediction for the evolution of the *BS*.

A task that could not be accomplished with the chosen setup is the in-situ characterization of the droplets accumulated at the channel centerline. This was due to the limitations in the optical resolution of the experimental setup that did not allow a characterization of the evolution of the stream of droplets, i.e. potential coalescence events. As an alternative, we collected a sample from the outlet of the microfluidic device and observed micron-sized DVB droplets through a microscope. These droplets are apparently much bigger than those which form in situ. It would be desirable to observe the evolution of the droplet size distribution inside the microchannel, However, to accomplish that an imaging system with a larger numerical aperture is needed.

**Acknowledgement**

R. Hajian acknowledges the Ministry of Science, Research and Technology of the I.R. Iran for funding under grant number 89100017. He also appreciates Dorothea Paulssen for her helpful hints on the "ouzo effect", and Tobias Baier for his effective discussions.